\documentclass[english,dvips,12pt]{iopart}
\usepackage{babel}
\usepackage{xcolor}
\usepackage{graphicx}
\usepackage[figlist]{endfloat}
\usepackage{textcomp}
\usepackage{amssymb}
\usepackage{amsopn}
\usepackage[per=slash,tophrase=\ensuremath{\div},noload=abbr]{siunitx}
\usepackage{cite}
\usepackage{url}

\DeclareMathOperator{\varRe}{Re}

\newcommand{\ju}{\ensuremath{\mathrm{j}}}
\newcommand{\eqdef}{\ensuremath{\stackrel{\mathrm{def}}{=}}}
\newcommand{\pa}{\ensuremath{\mathbin{||}}}
\renewcommand{\phi}{\varphi}
\renewcommand{\Re}{\varRe}

\newcommand{\LTspice}{LTspice IV\textsuperscript{\textregistered}}
\newcommand{\inrim}{INRIM}

\begin{document}
\bibliographystyle{unsrt}
\title[Systematic errors in JNTs]{\textbf{Systematic errors in the correlation method for Johnson noise thermometry: residual correlations due to amplifiers}}
\author{L Callegaro\textsuperscript{1}, M Pisani\textsuperscript{1} and M Ortolano\textsuperscript{1,2}}
\address{\textsuperscript{1} \inrim\ --- Istituto Nazionale di Ricerca Metrologica
Strada delle Cacce, 91, 10135 Torino, Italy}
\address{\textsuperscript{2} Dipartimento di Elettronica, Politecnico di Torino, Corso Duca degli Abruzzi, 24, 10129 Torino, Italy}
\eads{\mailto{l.callegaro@inrim.it}, \mailto{massimo.ortolano@polito.it}}

\begin{abstract}
Johnson noise thermometers (JNT) measure the equilibrium electrical noise, proportional to thermodynamic temperature, of a sensing resistor. In the \emph{correlation method}, the same resistor is connected to two amplifiers and a correlation of their outputs is performed, in order to reject amplifiers' noise. Such rejection is not perfect: the residual correlation gives a systematic error in the JNT reading. In order to put an upper limit, or to achieve a correction, for such error, a careful electrical modelling of the amplifiers and connections must be performed. Standard numerical simulation tools are inadequate for such modelling. In literature, evaluations have been performed by the painstaking solving of analytical modelling. We propose an evaluation procedure for the JNT error due to residual correlations which blends analytical and numerical approaches, with the benefits of both: a rigorous and accurate circuit noise modelling, and a fast and flexible evaluation with an user-friendly commercial tool. The method is applied to a simple but very effective ultralow-noise amplifier employed in a working JNT.
\end{abstract}
\maketitle

\section{Introduction}
A Johnson noise thermometer (JNT) determines the thermodynamic temperature $T$ of a sensing resistor $R$ by measuring its equilibrium voltage noise, given by the Johnson-Nyquist relation\cite{Nyquist:1928} $S_v(f) =4 k_\textup{B}T R$, where $S_v(f)$ is the noise voltage spectral density function and $k_\textup{B}$ is the Boltzmann constant\footnote{The low-frequency approximation of Johnson-Nyquist relation is reported. The expression is accurate to one part in \num{e7} at room temperature and frequency below \SI{1}{\mega\hertz}.}.

Johnson noise thermometry  has been employed successfully to measure temperatures in a wide range (from below \SI{100}{\milli\kelvin} to \SI{1800}{\celsius} \cite{Stenzel:2003}) and to determine temperature ratios between fixed points of the International Temperature Scale ITS-90 (see \cite{White:1996} for a review). In the framework of a redefinition of the \foreignlanguage{french}{Syst\`eme international d'unit\'es} (SI) in terms of fundamental constants, the use of JNTs performing measurements traceable to the representation of the volt given by the Josephson effect has been proposed as a method towards new determinations of $k_\textup{B}$ \cite{Fellmuth:2006,Benz:2009,Callegaro:2009}.

Most JNTs employ room-temperature electronics with junction field-effect transistor (JFET) front-end amplifiers. Such amplifiers have an equivalent input voltage noise having a spectral density of the same order of magnitude of Johnson noise.

The use of \emph{correlation} permits to reject to a large extent amplifiers' voltage noise (see \cite[par.\ 6.4]{White:1996} for a review). The resistor noise is amplified using two identical amplifiers, and only the correlated part of the amplified signals is considered for the measurement. Thanks to advances in mixed-signal electronics and real-time processing, digital implementation of correlators is preferred in more recent JNTs.

The rejection of amplifiers' noise given by the correlation method is not perfect. Amplifiers' voltage and current noises are coupled through the sensing resistor and its wiring, and give rise to residual correlation terms, indistinguishable from that given by Johnson noise, therefore giving a systematic offset in the temperature reading.

The amount of residual correlation is essentially dependent on the design of the input stage of the amplifier and the choice of its components. Some contributions to this residual correlation can be determined experimentally, with extreme difficulty, by regressions on measurements  with different values of $R$ of the same temperature \cite{Storm:1986}; other contributions are beyond direct experimental assessment.

Therefore, an evaluation must be performed by electrical modelling. We'll see in section \ref{sec:unbalanced} that the modelling has to take into account not only all relevant voltage and current noises generated by different sources, but also the correlations among them. Unfortunately, standard packages for numerical circuit simulations, such as SPICE \cite{Spice}, do not take into account such correlations, and therefore cannot be employed directly.

An alternative approach is analytical modelling. Equations modelling all circuit elements of the amplifier, and their corresponding noise sources, are written explicitly and coupled to circuit network equations~\cite{Buckingham:1983,Motchenbacher:1993}. The resulting equations can be solved analytically.

The analytical method has been applied to error analysis in JNTs since long \cite{White:1984}; White and Zimmermann \cite{White:2000} have made a comparative analysis of some simple amplifier circuits employed in JNT. The analytical method is complex and heavy, and in practice cannot be applied except for the simplest circuits. Any modification in the circuit mesh require modifications in the model equations, which must be solved again from scratch.

In the following, we propose a new method which blends the approach of the analytical method with numerical simulation of circuits. The method permits a rigorous noise analysis, taking into account noise correlations, with the practical advantages of numerical simulations (which an amplifier designer would probably employ in any case). The method will be applied to a working amplifier, very simple but of practical interest, successfully employed in a working JNT under development~\cite{Callegaro:2009}.

The following symbols conventions are used throughout this article: signals are represented by lowercase symbols with lowercase subscripts, whereas their Fourier transforms are represented by the corresponding uppercase symbols with the same lowercase subscripts (we will also consider ``Fourier transforms'' of random signals: for a more thorough discussion see \cite[ch.~3]{Buckingham:1983}); if $x(t)$ and $y(t)$ are both random signals, $S_x(f)$ and $S_y(f)$ represent their \emph{one-sided} spectral density functions ($f$ is the Fourier frequency) and $S_{xy}(f)$ represents their cross-spectral density function; for complex quantities, a star denotes complex conjugation and the operator $\Re$ takes the real part.

\section{Error analysis in unbalanced circuits}
\label{sec:unbalanced}

\subsection{Error equation}
In unbalanced measurements, a single resistor $R$ generates the noise voltage $v(t)$ which is then amplified by two amplifiers with unbalanced inputs and equal gains $A_v(f)$. The amplifiers' output signals $v_\textup{o1}(t)$ and $v_\textup{o2}(t)$ are then digitized and the cross-spectral density function between the two channels is estimated.

\begin{figure}
  \centering
  \includegraphics[clip=]{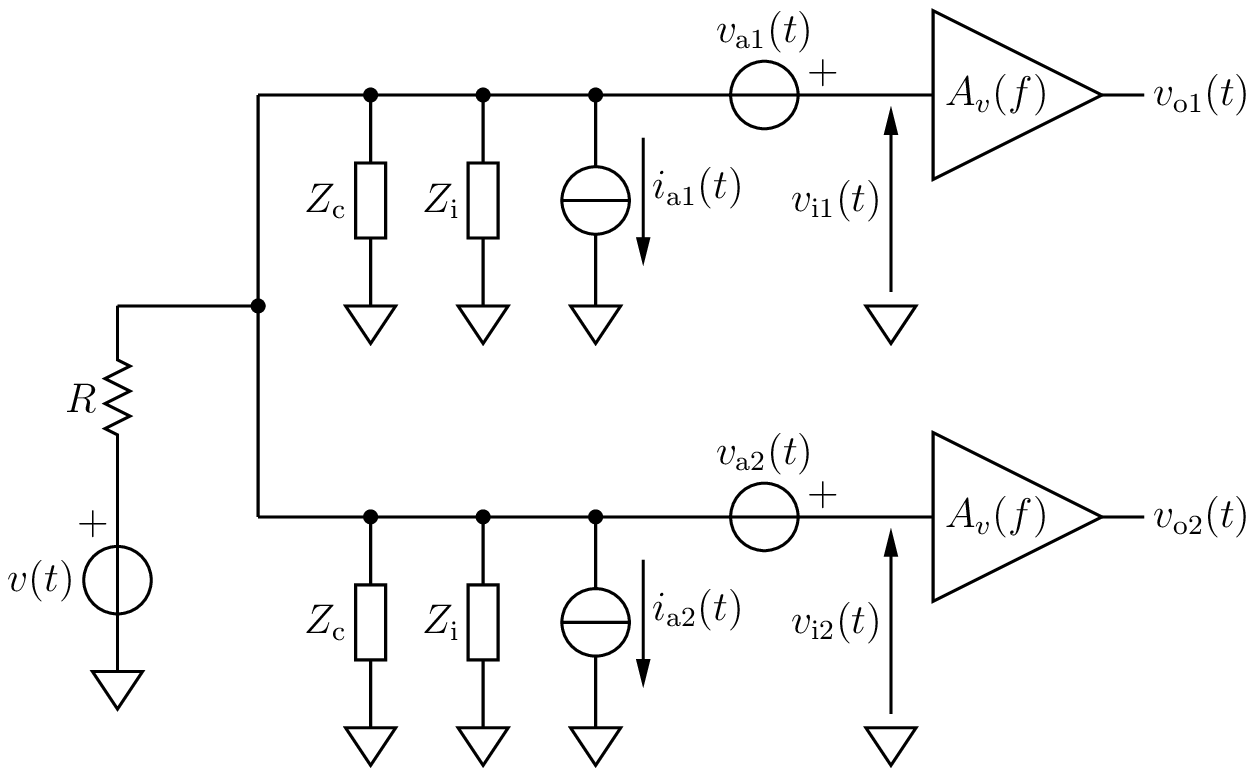}
  \caption{Equivalent circuit used in the analysis of unbalanced measurements.}\label{fig:circuit:unbalanced}
\end{figure}

The equivalent circuit for this kind of measurement is shown in figure~\ref{fig:circuit:unbalanced}. For simplicity, we will assume that the two channels have identical properties: where necessary, the given results can be easily generalized to consider unequal channels. The interconnection cables are modelled by the two impedances $Z_\textup{c}(f)$, while $Z_\textup{i}(f)$ represents the input impedance of the amplifiers. Voltage sources $v_\textup{a1}(t)$ and $v_\textup{a2}(t)$ represent the equivalent noise voltage of the amplifiers; current sources $i_\textup{a1}(t)$ and $i_\textup{a2}(t)$ represent the input \emph{short-circuit} noise current of the amplifiers. We will further assume that noise sources belonging to different amplifiers are uncorrelated: with this assumption, only the cross spectral densities $S_{v_\textup{a1}i_\textup{a1}}(f) = S_{v_\textup{a2}i_\textup{a2}}(f) = S_{v_\textup{a}i_\textup{a}}(f)$ need to be considered.

Let $\mathcal{T} = k_\textup{B}T$ and
\numparts
\begin{eqnarray}
  A(f) &= \frac{V_\textup{i1}(f)}{V(f)} = \frac{V_\textup{i2}(f)}{V(f)} \\
  &= \frac{1}{1+2R\left[1/Z_\textup{i}(f)+1/Z_\textup{c}(f)\right]}.
\end{eqnarray}
\endnumparts
If we assume the product $A(f)A_v(f)$ to be known from gain calibration --- which is typically the case with \emph{absolute} JNTs~\cite{Nam:2005}, then it can be shown that the relative measurement error of $\mathcal{T}$ due to the amplifiers is given by
\begin{equation}\label{eq:error:unbalanced}
  \frac{\Delta \mathcal{T}}{\mathcal{T}} = \frac{1}{2\mathcal{T}}\left\{R S_{i_\textup{a}}(f) - \frac{\Re\left[A(f)S^*_{v_\textup{a}i_\textup{a}}(f)\right]}{|A(f)|^2}\right\},
\end{equation}
The above equation is equivalent to~(6) of \cite{Storm:1989} and to~(4) of \cite{White:2000}. Generally, the error estimated from~\eref{eq:error:unbalanced} can be frequency-dependent: in such case, an average value in the frequency range of interest should be determined.

\subsection{Estimation of the spectral density functions $S_{i_\textup{a}}(f)$ and $S_{v_\textup{a}i_\textup{a}}(f)$}
Although it is conceivable to measure $S_{i_\textup{a}}(f)$ and $S_{v_\textup{a}i_\textup{a}}(f)$, in practice such operation is difficult and cannot be done with sufficient accuracy over the whole frequency range of interest. Therefore, some kind of analysis should be employed to evaluate $S_{i_\textup{a}}(f)$ and $S_{v_\textup{a}i_\textup{a}}(f)$. In~\cite{White:2000}, the analysis has been done with an analytical method; however, this approach is cumbersome and error-prone, and could only be employed with the simplest circuits. To circumvent these problems, we propose a method based on numerical simulations which is applicable to more complex circuits.

Consider one of the two amplifiers and let $x_j(t)$, which can be either a voltage or a current, be the $j$th noise source internal to the amplifier. Assuming the input short-circuited, the equivalent input noise voltage $v_\textup{a}(t)$ and the input short-circuit noise current $i_\textup{a}(t)$ of the amplifier can be written as a superposition of the noises generated by each source $x_j(t)$, i.e.
\numparts
\begin{eqnarray}
  V_\textup{a}(f) \eqdef \frac{V_\textup{o}(f)}{A_v(f)} = \sum_j Q_j(f)X_j(f) \label{eq:noisesuma} \\
  I_\textup{a}(f) = \sum_j P_j(f)X_j(f) \label{eq:noisesumb}
\end{eqnarray}
\endnumparts
The network functions
\numparts
\begin{eqnarray}
  Q_j(f) = \frac{V_{\textup{a}}(f)}{X_j(f)} = \frac{1}{A_v(f)}\frac{V_{\textup{o}}(f)}{X_j(f)} \\
  P_j(f) = \frac{I_{\textup{a}}(f)}{X_j(f)}
\end{eqnarray}
\endnumparts
can be approximately estimated by means of an analog circuit simulation program\footnote{In this work, we have used \LTspice\ from Linear Technology Corporation~\cite{LTspice}.}. In particular, such programs can perform small-signal analysis to determine directly the ratios $V_{\textup{o}}(f)/X_j(f)$, $I_{\textup{a}}(f)/X_j(f)$ and $A_v(f)$ and the input impedance $Z_\textup{i}(f)$.

Using~\eref{eq:noisesuma}--\eref{eq:noisesumb}, it is not difficult to show that the searched spectral densities can be written as ($S_{v_\textup{a}}(f)$ is given for completeness, although it does not enter explicitly in~\eref{eq:error:unbalanced})
\numparts
\begin{eqnarray}
  S_{v_\textup{a}}(f) = \sum_{j,k} Q_j(f)Q^*_k(f)S_{x_jx_k}(f) \label{eq:sdf-a:unbalanced} \\
  S_{i_\textup{a}}(f) = \sum_{j,k} P_j(f)P^*_k(f)S_{x_jx_k}(f) \\
  S_{v_\textup{a}i_\textup{a}}(f) = \sum_{j,k} Q_j(f)P^*_k(f)S_{x_jx_k}(f) \label{eq:sdf-c:unbalanced}
\end{eqnarray}
\endnumparts
where $S_{x_jx_k}(f)$ is the cross-spectral density function between the noise sources $x_j(t)$ and $x_k(t)$, with the convention that, if $j=k$, $S_{x_jx_j}(f)$ represents the spectral density function of $x_j(t)$. The assignment of the functions $S_{x_jx_k}(f)$ must be based on the known noise models of the devices composing the specific amplifier under examination. Actually, most of the $S_{x_jx_k}(f)$ with $j\neq k$ can be considered identically zero.

Although not easy to evaluate, the uncertainties of the estimations given by equations~\eref{eq:sdf-a:unbalanced}--\eref{eq:sdf-c:unbalanced} are related to the uncertainties of the various parameters on which the functions $P_j(f)$, $Q_j(f)$ and $S_{x_jx_k}(f)$ depend, and would be the same obtainable with the analytic method.

\section{Error analysis in balanced circuits}
\label{sec:balanced}
\begin{figure}
  \centering
  \includegraphics[clip=]{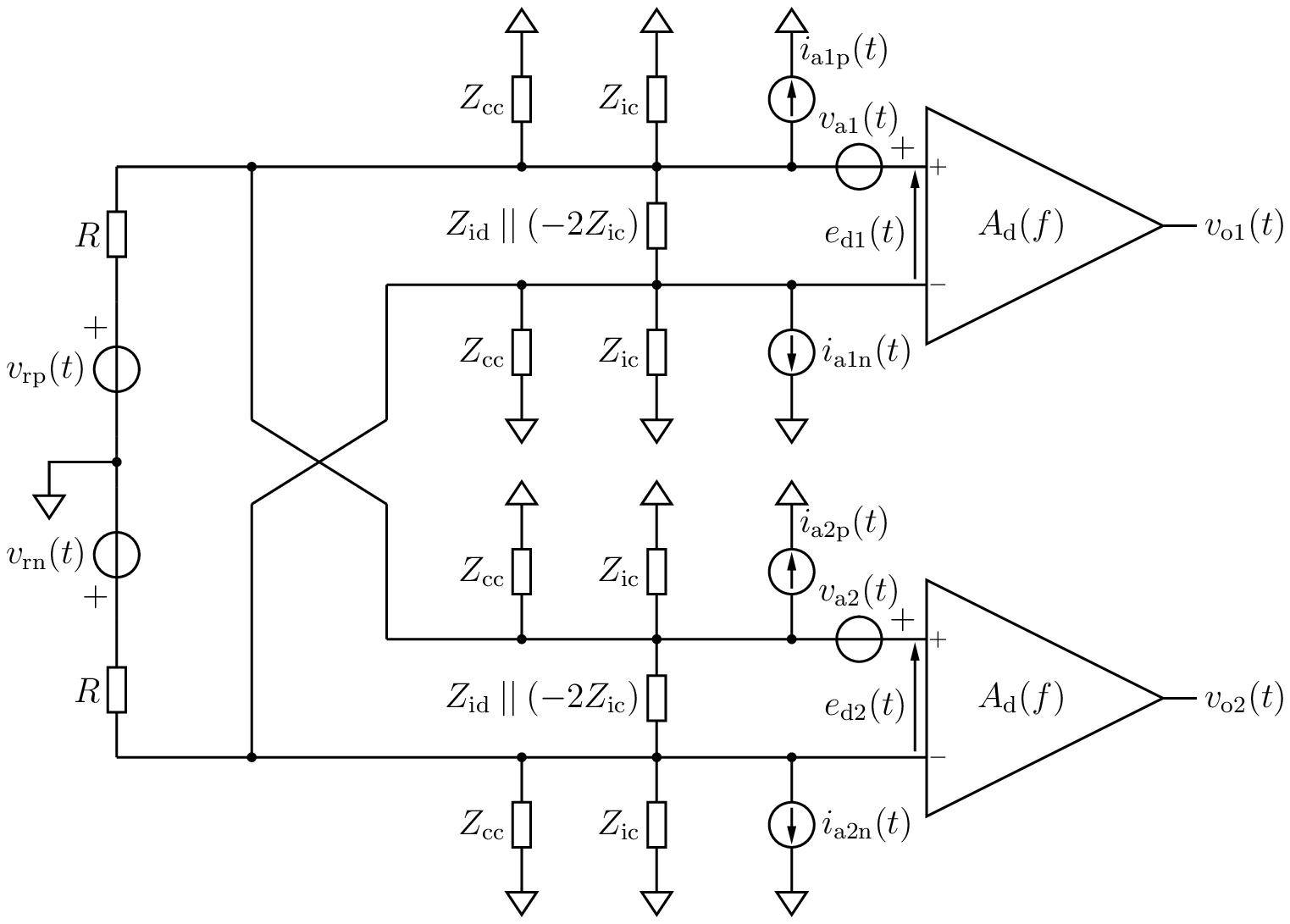}
  \caption{Equivalent circuit used in the analysis of balanced measurements: $Z_\textup{id}$ represents the differential-mode input impedance of the amplifiers; $Z_\textup{ic}$ represents the common-mode input impedance; and $Z_\textup{id}\pa (-2Z_\textup{ic}) = -2Z_\textup{ic} Z_\textup{id}/(Z_\textup{id}-2Z_\textup{ic})$ (see \cite{Gray:2001} for details).}\label{fig:circuit:balanced}
\end{figure}

Typically, in order to improve the rejection to external disturbances, balanced measurements are preferred over unbalanced ones. The equivalent circuit for a balanced measurement is shown in figure~\ref{fig:circuit:balanced}: two instrumentation amplifiers with differential gain $A_\textup{d}(f)$ measure the differential noise voltage generated by two sensing resistors working at the same thermodynamic temperature $T$. Again, the amplifiers' output signals are digitized and the cross-spectral density function between the two channels is estimated. The symbols in figure~\ref{fig:circuit:balanced} have the following meaning: $v_\textup{rp}(t)$ and $v_\textup{rn}(t)$ are the noise voltages generated by the sensing resistors $R$; $Z_\textup{cc}(f)$ models the impedance of the interconnection cables; $Z_\textup{id}(f)$ and $Z_\textup{ic}(f)$ represent respectively the differential-mode and common-mode input impedances of the amplifiers \cite{Gray:2001}; $v_\textup{a1}(t)$ and $v_\textup{a2}(t)$ represent the equivalent noise voltage of the amplifiers; $i_\textup{a1p}(t)$, $i_\textup{a1n}(t)$, $i_\textup{a2p}(t)$ and $i_\textup{a2n}(t)$ represent the input short-circuit noise currents of the amplifiers.

Let
\numparts
\begin{eqnarray}
  A(f) &= \frac{E_\textup{d1}(f)}{V_\textup{rd}(f)} = \frac{E_\textup{d2}(f)}{V_\textup{rd}(f)} \\
  &=\frac{1}{1+2R\left[2/Z_\textup{id}(f)+1/Z_\textup{cc}(f)\right]}
\end{eqnarray}
\endnumparts
then, in this case, with the same assumptions made in \S\ref{sec:unbalanced}, the relative measurement error of $\mathcal{T}=k_\textup{B}T$ due to the amplifiers is given by
\begin{eqnarray}\label{eq:error:balanced}
  \frac{\Delta \mathcal{T}}{\mathcal{T}} = &\frac{1}{4\mathcal{T}}\left\{2R\left[S_{i_\textup{a}}(f) -\Re S_{i_\textup{ap}i_\textup{an}}(f)\right]\vphantom{\frac{\Re\left\{A(f)\left[S^*_{v_\textup{a}i_\textup{ap}}(f) - S^*_{v_\textup{a}i_\textup{an}}(f)\right]\right\}}{|A(f)|^2}}\right. \nonumber \\
  &\left.-\frac{\Re\left\{A(f)\left[S^*_{v_\textup{a}i_\textup{ap}}(f) - S^*_{v_\textup{a}i_\textup{an}}(f)\right]\right\}}{|A(f)|^2}\right\},
\end{eqnarray}
where $S_{i_\textup{a}}(f)$ is the spectral density function of $i_\textup{a1p}(t)$, $i_\textup{a1n}(t)$, $i_\textup{a2p}(t)$ and $i_\textup{a2p}(t)$, $S_{i_\textup{ap}i_\textup{an}}(f)$ is the cross-spectral density function between $i_\textup{a1p}(t)$ and $i_\textup{a1n}(t)$ (or between $i_\textup{a2p}(t)$ and $i_\textup{a2n}(t)$), $S_{v_\textup{a}i_\textup{ap}}(f)$ is the cross-spectral density function between $v_\textup{a1}(t)$ and $i_\textup{a1p}(t)$ (or between $v_\textup{a2}(t)$ and $i_\textup{a2p}(t)$), $S_{v_\textup{a}i_\textup{an}}(f)$ is the cross-spectral density function between $v_\textup{a1}(t)$ and $i_\textup{a1n}(t)$ (or between $v_\textup{a2}(t)$ and $i_\textup{a2n}(t)$).

The procedure outlined in \S\ref{sec:unbalanced} can be generalized to yield the above defined spectral densities:
\numparts
\begin{eqnarray}
  S_{v_\textup{a}}(f) = \sum_{j,k} Q_j(f)Q^*_{\textup{n}k}(f)S_{x_jx_k}(f) \label{eq:sdf-a:balanced} \\
  S_{i_\textup{a}}(f) = \sum_{j,k} P_{\textup{p}j}(f)P^*_{\textup{p}k}(f)S_{x_jx_k}(f) = \sum_{j,k} P_{\textup{n}j}(f)P^*_{\textup{n}k}(f)S_{x_jx_k}(f)   \\
  S_{i_\textup{ap}i_\textup{an}}(f) = \sum_{j,k} P_{\textup{p}j}(f)P^*_{\textup{n}k}(f)S_{x_jx_k}(f) \\
  S_{v_\textup{a}i_\textup{ap}}(f) = \sum_{j,k} Q_j(f)P^*_{\textup{p}k}(f)S_{x_jx_k}(f) \\
  S_{v_\textup{a}i_\textup{an}}(f) = \sum_{j,k} Q_j(f)P^*_{\textup{n}k}(f)S_{x_jx_k}(f)
\end{eqnarray}
\endnumparts
with
\numparts
\begin{eqnarray}
  Q_j(f) = \frac{1}{A_\textup{d}(f)}\frac{V_{\textup{o}}(f)}{X_j(f)} \\
  P_{\textup{p}j}(f) = \frac{I_{\textup{ap}}(f)}{X_j(f)} \\
  P_{\textup{n}j}(f) = \frac{I_{\textup{an}}(f)}{X_j(f)} \label{eq:nf-c:balanced}.
\end{eqnarray}
\endnumparts
Also in this case, an analog circuit simulation program can be used to estimate the network functions $Q_j(f)$, $P_{\textup{p}j}(f)$ and $P_{\textup{n}j}(f)$.

\section{Example of application}
\begin{figure}
  \centering
  \includegraphics[clip=]{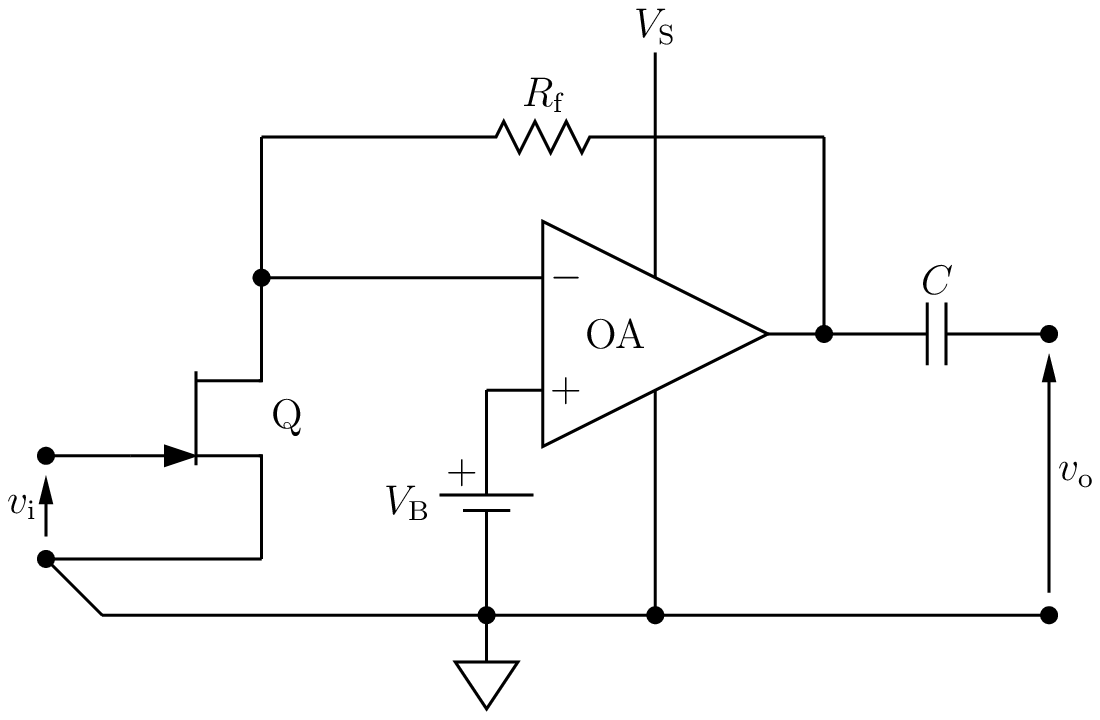}
  \caption{The amplifier schematics. See text for an explanation of the symbols.}
  \label{fig:schema}
\end{figure}

The \inrim\ JNT~\cite{Callegaro:2009} employs two amplifiers with the circuit of figure~\ref{fig:schema}. Despite its simplicity, it has a voltage noise below \SI{1}{\nano\volt\per\sqrt{\hertz}} at audio frequencies, which is much lower than the best JFET integrated circuit amplifiers, and comparable to ultralow-noise discrete amplifiers (see e.g.\ \cite{Klein:1979,Jefferts:1989,Neri:1991,Howard:1999,Ferrari:2002,Zhang:2006}) which are however more complex. The value of the sensing resistor is \SI{1}{\kilo\ohm} and the interconnection cables have a capacitance of about \SI{80}{\pico\farad}.

\subsection{Description}
Looking at figure~\ref{fig:schema} we see that the JFET Q, directly connected to the input voltage $v_\textup{i}$, works in a common-source configuration with a gate-source bias voltage\footnote{A drawback of such configuration is the limited dynamic range, a few tens of \milli\volt\ at the input, before distortion occurs; however, in a JNT the integrated signal is of a few \micro\volt.} $V_\textup{GS} \approx \SI{0}{\volt}$.


The operational amplifier OA works as a transresistance amplifier with gain $R_\textup{f}$, and biases Q; the transresistance configuration eliminates the Miller effect \cite[ch.\ 7]{Gray:2001}, thus enhancing the bandwidth. The drain-source voltage $V_\textup{DS}$ is set by OA at the voltage $V_\textup{B}\approx \SI{9}{\volt}$ of the polarization battery,
which has an extremely low noise \cite{Boggs:1995,Hassibi:2004}.

OA works in a single-supply configuration; its output $v_\text{o}$ is ac-coupled through capacitor $C$, and can be further amplified by additional stages if necessary.

The overall low-frequency gain $A_{v0} = g_\textup{m}R_\textup{f}$ of the amplifier depends on $R_\textup{f}$ and on the transconductance $g_\textup{m}$ of the JFET; the bandwidth of the amplifier depends on the JFET capacitances and on the gain-bandwidth product of OA; such parameters can have significant deviations from sample to sample and have a strong temperature dependence. In JNT applications, where periodic calibration \cite{White:1996} of $A_v(f)$ is performed with a reference signal, this does not constitute a problem.

Typical supply voltage is $V_\textup{S} \approx \SI{24}{\volt}$ from an unregulated battery; the quiescent current is about \SI{15}{\milli\ampere}.

\subsection{Construction and performance}
\label{sec:construction}
A prototype has been assembled with a 2SK170 JFET (Toshiba), an OP27 (various suppliers) for OA and $R_\textup{f}=$\SI{1}{\kilo\ohm}. The gate current $I_\textup{g}$ is \SIrange{2}{3}{\pico\ampere} (measured with a Keithley mod.\ 6430 current meter), which gives a current shot noise less than \SI{1}{\femto\ampere\per\sqrt{\hertz}}.

\begin{figure}
  \centering
  \includegraphics[width=\textwidth,clip]{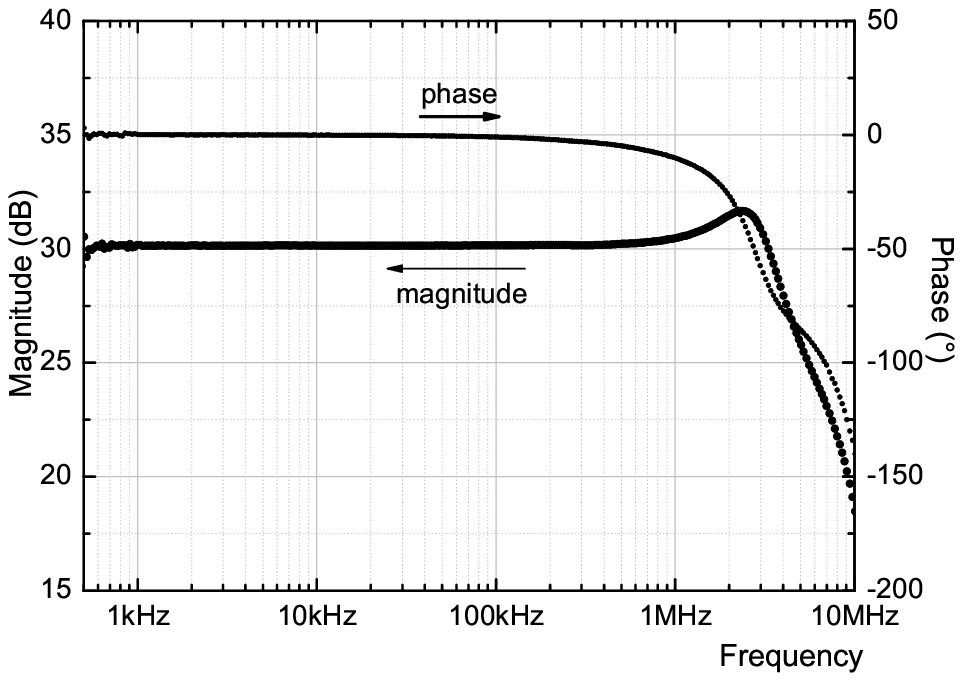}
  \caption{The transfer function of the amplifier, in magnitude and phase representation.}
  \label{fig:transfunc}
\end{figure}

The transfer function of the amplifier is shown in figure~\ref{fig:transfunc}. It has been measured with a network
analyzer (Agilent Tech.\ mod.\ 4395A), injecting the signal with a resistive divider (\SI{50}{\ohm}-\SI{0.5}{\ohm}). A \SI{3}{\deci\bel}-bandwidth of about \SI{4}{\mega\hertz} can be estimated, with a gain flatness better than \SI{1}{\deci\bel} up to \SI{1}{\mega\hertz}.

\begin{figure}
  \centering
  \includegraphics[width=\textwidth,clip]{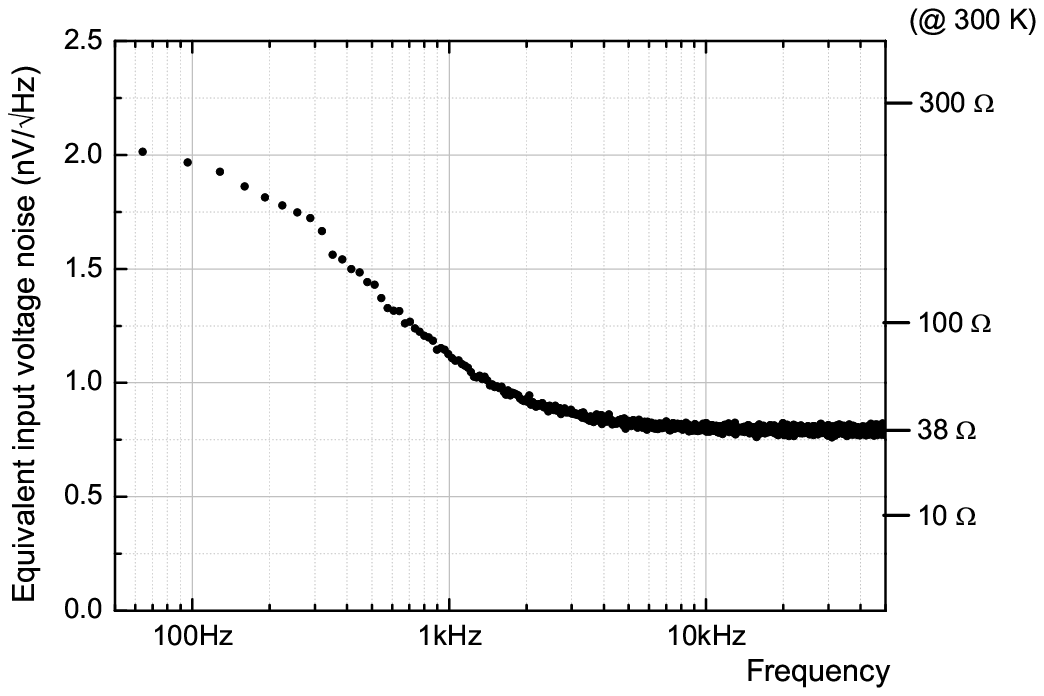}
  \caption{Equivalent input voltage noise of the amplifier.}
  \label{fig:noisefloor}
\end{figure}

The square root of $S_{v_\textup{a}}(f)$ is shown in figure~\ref{fig:noisefloor}. It has been measured with a two-channel signal analyzer (Agilent Tech.\ mod.\ 35670A) by connecting $v_\textup{o}$ to both channels and performing a cross-correlation measurement in order to reject the analyzer noise. The noise floor is about \SI{0.8}{\nano\volt\per\sqrt{\hertz}}, corresponding to the Johnson noise of a $\approx$\SI{38}{\ohm} resistor at
\SI{300}{\kelvin}. We do not consider the noise spectrum for frequencies below \SI{1}{\kilo\hertz}, and hence we neglect the effect of low-frequency noises such as generation-recombination and flicker noises, because the operating frequencies of JNTs are typically well above \SI{1}{\kilo\hertz}.

\subsection{Results}
\begin{figure}
  \centering
  \includegraphics[clip=]{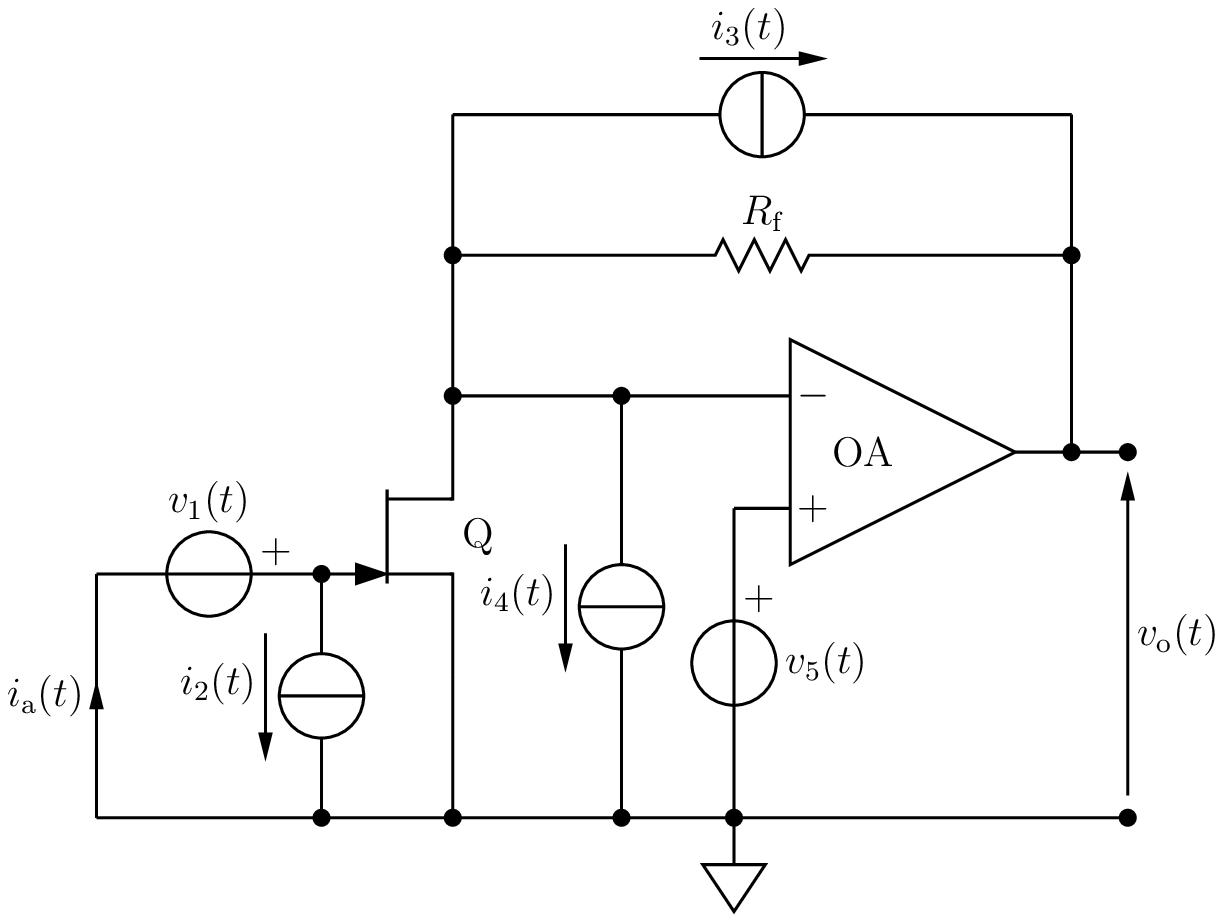}
  \caption{Small-signal equivalent circuit of the amplifier of figure~\ref{fig:schema} with noise sources.}\label{fig:jfetnoise}
\end{figure}
Figure~\ref{fig:jfetnoise} shows the small-signal equivalent circuit of the amplifiers with the five noise sources considered in the error analysis: $v_1(t)$ and $i_2(t)$ represent respectively the equivalent voltage and current input noise generators of the JFET Q; $i_3(t)$ represents the thermal noise of $R_\textup{f}$; $i_4(t)$ and $v_5(t)$ represent respectively the equivalent current and voltage input noise generators of the operational amplifier OA. $v_1(t)$ and $i_2(t)$ have been assumed with spectral density functions~\cite[ch.~5]{Buckingham:1983} ($C_\textup{gs}$ is the gate-source capacitance of the JFET)
\numparts
\begin{eqnarray}
  S_{v_1}(f) \approx \frac{8k_\textup{B}T}{3g_\textup{m}} \\
  S_{i_2}(f) \approx 2qI_\textup{g}+\frac{(2\pi f C_\textup{gs})^2}{g_\textup{m}}k_\textup{B}T
\end{eqnarray}
and cross-spectral density function~\cite[ch.~5]{Buckingham:1983}
\begin{equation}
  S_{v_1 i_2}(f)\approx 0.4\ju\, \left[S_{v_1}(f)S_{i_2}(f)\right]^{1/2}.
\end{equation}
\endnumparts
$i_3(t)$ has been assumed with spectral density function $S_{i_3}(f) = 4k_\textup{B}T/R_\textup{f}$, while the spectral densities of $i_4(t)$ and $v_5(t)$ have been inferred from the operational amplifier's data sheet. Apart from $v_1(t)$ and $i_2(t)$, all other generators have been considered uncorrelated.

\begin{figure}
  \centering
  \includegraphics[clip=]{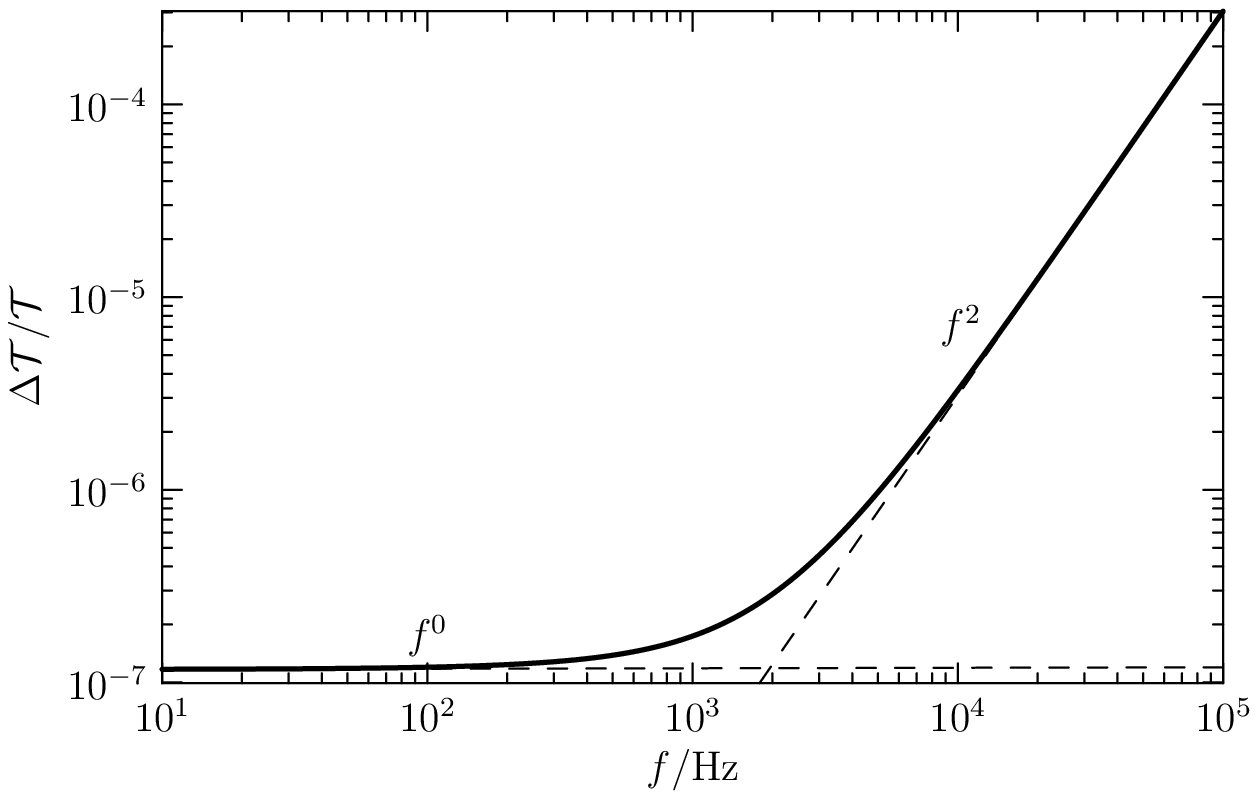}
  \caption{Expected temperature measurement error as a function of frequency for the amplifier of figure~\ref{fig:schema} (more details in text).}\label{fig:error}
\end{figure}

The expected temperature measurement error resulting from the simulation is shown in figure~\ref{fig:error} as a function of frequency. For low frequencies the error tends asymptotically to a constant: here, the main error source is the gate current shot noise of the JFET. At high frequencies the error tends to increase as $f^2$: this behaviour depends mainly on: i) $S_{i_2}(f)$ which increases as $f^2$, ii) $v_1(t)$ which contributes to $i_\textup{a}(t)$ through the input impedance of the amplifier, and iii) $v_5(t)$ which contributes to $i_\textup{a}(t)$ through the gate-source capacitance of the JFET.

From figure~\ref{fig:error}, the error can be approximated by
\begin{equation}
\frac{\Delta \mathcal{T}}{\mathcal{T}}\approx \num{1.2e-7}+\num{3.04e-14}(f/\hertz)^2
\end{equation}
and the expected average error over the present operating frequency range of the \inrim\ JNT (\SIrange{3}{7}{\kilo\hertz}) is about \num{9.2e-7}.

\section{Conclusions}
We have described a method for the evaluation of the systematic error of the temperature reading of a correlation JNT caused by residual terms in the correlation spectrum of the amplified Johnson noise. Although a general expression for such error can be given (equation~\eref{eq:error:balanced}), its terms are dependent on circuit components and topology. The method consists in the identification of noise sources of the components employed in the amplifiers' network; in the use of a commercial circuit simulation software, with its user-friendly interface, to numerically estimate a number of transfer functions for such noise signals; and in a numerical calculation (equations~\eref{eq:sdf-a:balanced}--\eref{eq:nf-c:balanced}) easy to implement on any platform. Versions of the model for both single-ended and differential input amplifiers are given.

The paper shows the application of the method to a simple but effective single-ended open-loop amplifier of practical interest. However, the method can be easily applied to much more complex amplifier networks for which an analytical approach would be unfeasible.

Future developments will be devoted to the application of the method to other JNT amplifiers, either already working or under development, having both open- and closed-loop topologies.

\ack
The authors are indebted with their colleagues V.\ D'Elia and A.\ Pollarolo for their help in the experimental characterization of the amplifiers.

\section*{References}
\bibliography{jnt}
\end{document}